# Temperature dependent photoluminescence of single CdS nanowires

Thang Ba Hoang, L.V. Titova*, H.E. Jackson, and L.M. Smith
*Department of Physics, University of Cincinnati, Cincinnati, Ohio 45221*

J. M. Yarrison-Rice
*Department of Physics, Miami University, Oxford, Ohio 45056*

J.L. Lensch and L.J. Lauhon
*Department of Materials Science and Engineering, Northwestern University, Evanston, Illinois 60208*

Temperature dependent photoluminescence (PL) is used to study the electronic properties of single CdS nanowires. At low temperatures, both near-band edge (NBE) photoluminescence (PL) and spatially-localized defect-related PL are observed in many nanowires. The intensity of the defect states is a sensitive tool to judge the character and structural uniformity of nanowires. As the temperature is raised, the defect states rapidly quench at varying rates leaving the NBE PL which dominates up to room temperature. All PL lines from nanowires follow closely the temperature-dependent band edge, similar to that observed in bulk CdS.

*Author to whom the correspondence should be addressed. Electronic mail:

ltitova@physics.uc.edu

Recent advances in fabrication of CdS nanowires by the vapor-liquid-solid (VLS) growth technique[1] have resulted in the development of nanoscale optical devices including nanowire-based photodetectors,[2] waveguides,[3] and lasers.[4,5] Understanding the nature of nanowire defects and how to control them is particularly important for the future development of nanowire-based structures and devices. Unlike bulk semiconductors, however, nanowires have a large surface to volume ratio and so nearly every physical property of nanowires is highly sensitive to surface quality and wire morphology. One should therefore expect that nanowire defects as well should be strongly affected by the local nanowire structure and geometry. Single nanowire spectroscopy provides an essential component for understanding the nature of defects in these nanostructures.

In this work, we use micro-photoluminescence (PL) spectroscopy to study the temperature-dependent photoluminescence from several *single* CdS nanowires from 5 K to 295 K. We find that at room temperature, emission from *all* nanowires is dominated by PL from relatively broad excitonic emission near the CdS band edge. Room-temperature PL is thus surprisingly insensitive to structural differences between individual nanowires. However, at low temperatures, the internal structure of the nanowires produces two bands of PL related to (1) exciton emission from near the CdS band edge, and (2) defect emission at lower energies. Thus, low temperature PL not only provides valuable information about the electronic structure of the nanowires, but also provides ready insight into their structural quality. This paper looks in detail at both near band-edge and defect emission from several different single CdS nanowires as a function of temperature in order to draw some conclusions about the nature of CdS nanowire defects.

The CdS nanowires were synthesized using techniques described previously[1]. For single nanowire measurements, the wires were dispersed from the growth substrate into methanol solution and deposited onto a silicon substrate, resulting in a dilute array of nanowires with diameters ranging from 50 to 200 nm and lengths of 10 to 15 µm. The substrate with the dilute nanowire array was placed into a continuous flow helium cryostat where the temperature can be varied from 4 - 300 K.

PL from single nanowires was obtained through slit-confocal micro-photoluminescence as detailed in Reference 6.[6] Several individual isolated nanowires were identified and excited by 2.5 mW of the 458 nm line of an Ar+ laser. The laser beam was defocused to a 20 µm spot diameter in order to uniformly illuminate the entire nanowire of interest. A 50X/0.5NA long working length microscope objective was used to collect the PL emission. A 250X magnified image of the nanowire was oriented along the entrance slit of the spectrometer using a Dove prism in order to collect PL emission from the entire nanowire at one time. The PL was dispersed by a DILOR triple spectrometer working in subtractive mode, and detected by a 2000 x 800 pixel liquid nitrogen-cooled CCD detector.

We have studied ten different CdS nanowires with diameters ranging from 50 to 200 nm. Since the sizes of the nanowires are significantly larger than the Bohr exciton radius in CdS (~2.8 nm), no quantum confinement effects are expected. As we reported earlier,[7] these nanowires display a wide range of morphologies from straight and uniform to quite irregular with bends, kinks, and irregular shaped clusters of material aggregating locally on the nanowire surface. Such morphological variations may originate either during the growth of the nanowires, or during the deposition of the nanowires onto the Si substrate.

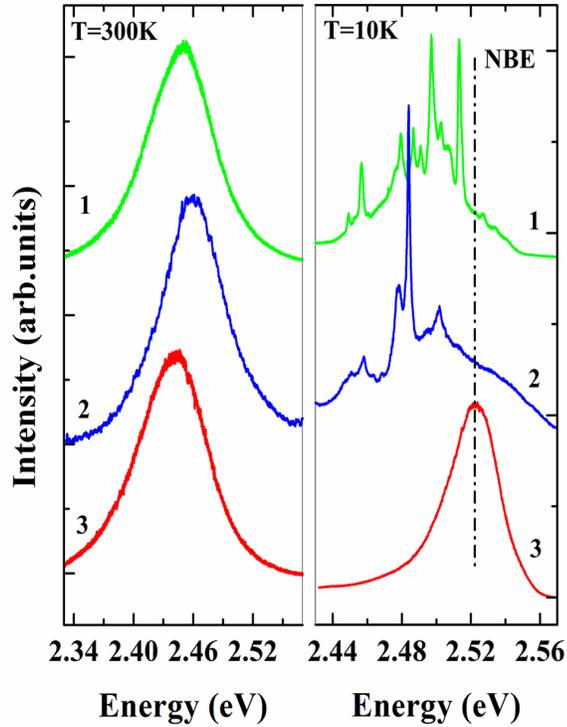

FIG. 1. (Color online) Room and low temperature PL spectra of 3 different wires 1, 2 and 3.

Figure 1 shows room temperature and low temperature PL spectra of three different wires, one of which is straight and uniform (wire 3), and two others that exhibit significant morphological irregularities (wires 1 and 2). At room temperature (Fig. 1(a)), PL spectra of all wires regardless of their morphology are alike and consist of a single broad (FWHM 30 - 70 meV) peak. Since the energy of this peak is below the bulk CdS band gap (2.51 eV), but in close proximity to the emission of bound exciton complexes in CdS (such as acceptor and donor bound excitons) (15), we label this peak "near band-edge (NBE) emission".

In contrast, the low temperature PL spectra (10 K, Fig. 1(b)) of these single nanowires differ significantly. The uniform wire (3) continues to emit a single peak at an energy of 2.52 eV which is comparable to the energy for excitonic emission from bulk CdS, while the PL spectra from the irregularly-shaped nanowires (1 and 2) exhibit a series of narrow lines 30 to 60 meV lower in energy than the broad NBE emission.  For these wires, the NBE emission appears only as a high energy shoulder to the PL emission band. Additional wires show behavior intermediate to these two sets of wires.  Correlation of the detailed spatially-resolved photoluminescence studies with AFM imaging[7] indicates that the narrow lines in the low temperature PL spectra are emitted at spatially distinct positions along the nanowire length and can be associated with the presence of morphological irregularities (kinks or lobes) in CdS nanowires.  Completely smooth and regular wires show only NBE emission.  The nature of these defects and the mechanisms of the carrier localization at the defect sites are still open questions.

In order to provide some insight into these defect-related states, we show in Fig. 2 the PL emission from wire 1 at a number of different temperatures from 5 K to 295 K.  All spectra have been normalized for clarity. At low temperatures (<20 K), the imperfection-related sharp lines are prominent, while the NBE emission appears only as a high energy shoulder.  As the temperature increases, the sharp lines exhibit a marked decrease in intensity starting around 30 K and are completely quenched by 80 - 90 K.  As the temperature increases further, a single NBE emission peak is observed.  The center of the NBE peak redshifts, as expected, above 90K, and persists to room temperature.

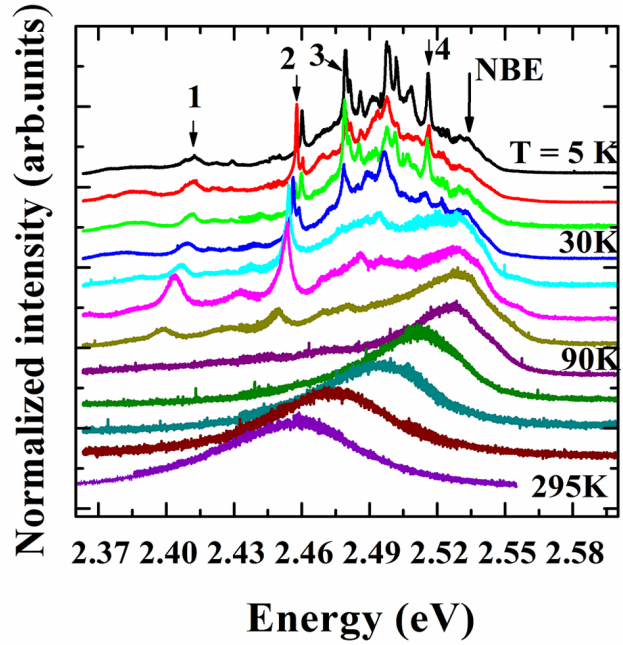

FIG. 2. (Color online) Temperature dependent PL of wire 1. Spectra have been normalized and offset for clarity.

The temperature dependence of the NBE peak energy, as well as the energies of several of the narrow lines from nanowire 1 are shown in Fig. 3. For comparison, we show the temperature dependence of the band-edge for bulk CdS single crystals as measured by Imada and collaborators.[8] The temperature dependence of both the bulk CdS band-edge and the nanowire NBE energy were fit to the extended phenomenological Varshni equation[9]

$$E_g = E_0 - \frac{\alpha T^4}{\beta + T^3}.$$

This form of the Varshni equation is consistent with the recent result by Cardona, Meyer and Thewalt, who used general thermodynamic arguments to show that the appropriate energy gap temperature dependence in the low temperature regime (as T goes to 0) should be $T^4$ rather than $T^2$.[10]

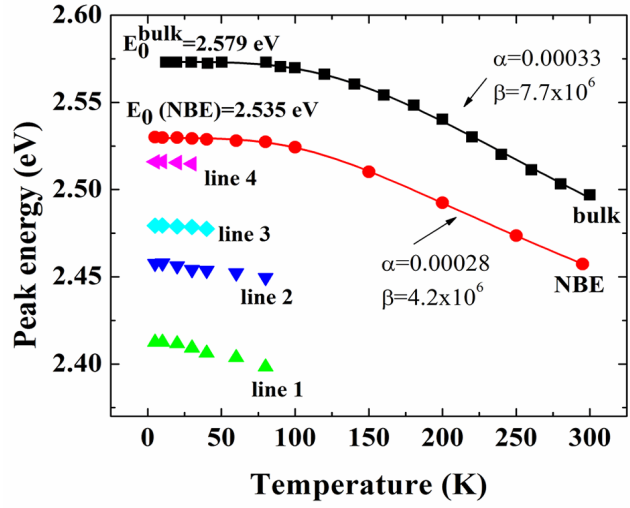

FIG. 3. (Color online) Energies of NBE as well as several localized state emissions as function of temperature. Temperature dependence of band gap of bulk CdS is shown for reference.

Good fits to the nanowire NBE data were obtained using the fitting parameters $E_0$=2.53eV, $\alpha$=0.00028, and $\beta$=4.2×10$^6$, while the bulk fitting parameters were $\alpha$=0.00033, and $\beta$=7.7×10$^6$  The slight variation between the nanowire and bulk Varshni parameters likely results from an intrinsic difference between nanowire and bulk crystal structure and the particular sensitivity of the nanowires to the surface properties and strain due to their small size and large surface-to-volume ratio.

The energies of the defect-related sharp lines (lines 1-4 in Fig. 3) follow the NBE peak closely as temperature increases until they are quenched.  This suggests that the narrow lines are not related to deep levels, and implies that the defect-related emission is excitonic in nature.  This conclusion is in agreement with previous time-resolved measurements on these nanowires, which showed that the recombination lifetimes of these states are 350 to 1000 ps, which are comparable to excitonic lifetimes measured in bulk CdS.[11,12]

To further study the rapid quenching of the sharp spectral features, we plot the integrated intensities of the PL as a function of temperature. The temperature dependence of the integrated intensities of PL emitted from the nanowire as a whole is plotted in Fig. 4(a) and shows that the overall PL efficiency decreases somewhat as the temperature increases from 10 K to 150 K, but it remains fairly constant from there up to room temperature. The integrated intensity of the defect-related PL lines is obtained through subtraction of the background and integration of the isolated peak (see inset to Fig. 4(b)). The results from this procedure are plotted for three spectral lines and shown in Fig. 4(b). The temperature dependent intensities of these three lines show unique temperature dependencies, and quench at different temperatures. However, all defect-related PL lines disappear by 90 K. One might conclude that excitons are only weakly bound to these defects by 4 to 8 meV, which is not consistent with the fact that these lines emit nearly 30 to 60 meV below the exciton complexes near the band edge. Clearly, some configuration of the defect, perhaps a charge state, is only stable at low temperature. Excitons are only localized at defects with this configuration. Moreover, recent 10 K time-resolved PL measurements of these states also show recombination lifetimes ranging from 300 to 1000 ps. All of these facts, taken together suggest that these lines result from exciton localization to defects, but with emission energies, quenching behavior and lifetimes which vary significantly, possibly due to varying local properties or structure of the nanowire near the defect site.

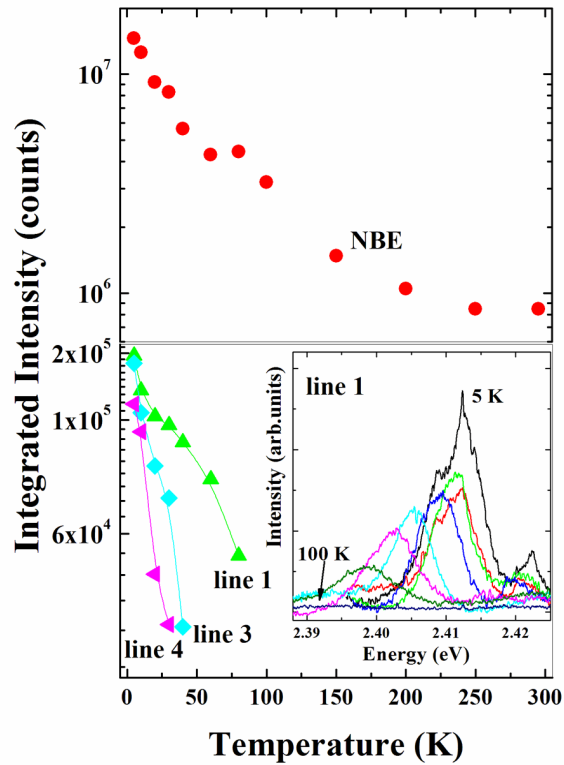

FIG. 4. (Color online) Intensities of NBE (a) as well as several localized state emissions (b) as function of temperature; temperature progression of one of the localized state emissions (line 1) as a

In conclusion, we have shown that the room temperature PL of single CdS nanowires is not sensitive to the structure of or defects within each nanowire. Indeed, both straight, uniform nanowires and nanowires with significant morphological irregularities emit a single peak near the band edge at room temperature. In strong contrast, low temperature PL emission is extremely sensitive to the nanowire morphology. We find that the low temperature (<20K) PL spectra of single nanowires exhibits two bands: one which is observed in all studies of nanowires regardless of morphological quality and is associated with bound excitons at the band edge, and the other which is associated with excitons localized to morphological defects along the nanowires. These defect-related states manifest themselves as narrow emission lines at a variety of lower energies, which track with the temperature dependence of the band edge, but exhibit unique quenching behaviors with temperature. Moreover, each defect line exhibits a unique recombination lifetime. This rather remarkable variability strongly suggests that nanowire defects are sensitive to the local properties or structure of the nanowire. All such defect lines disappear by 90 K, leaving the near band edge emission which persists up to room temperature. Because of the sensitivity to the nanowire surface and morphology, we have shown that low temperature PL measurements of single nanowires are absolutely essential to understanding the properties of defects in these structures.

**Acknowledgements:**

TBH and LVT are both equally responsible for this paper. This work was supported by the National Science Foundation through grants DMR 0071797 and 0216374, the Petroleum Research Fund of the American Chemical Society, University of Cincinnati and Northwestern University. J.L.L. acknowledges the support of a National Science Foundation Graduate Research Fellowship.